\documentclass[aps,prl,twocolumn,floatfix,showpacs,superscriptaddress]{revtex4-1}
\usepackage{amsmath,amsfonts,amssymb,bm}
\usepackage{graphicx}
\usepackage{color}
\usepackage{subfigure}
\usepackage{multirow}
\usepackage{textcomp}
\usepackage{slashed}
\usepackage[utf8]{inputenc}

\definecolor{purple}{rgb}{0.5,0,0.5}
\definecolor{blue}{rgb}{0.0,0,1.0}

\usepackage[colorlinks=true, pdfstartview=FitV, linkcolor=purple, citecolor= purple, urlcolor=blue]{hyperref}
\begin{document}


\title{Can the hyperfine mass splitting formula in heavy quarkonia be applied to the $B_c$ system?}

\author{Lei Chang}\email{leichang@nankai.edu.cn}
\affiliation{School of Physics, Nankai University, Tianjin 300071, China}

\author{Muyang Chen}\email{muyang@nankai.edu.cn}
\affiliation{School of Physics, Nankai University, Tianjin 300071, China}

\author{Xue-qian Li}\email{lixq@nankai.edu.cn}
\affiliation{School of Physics, Nankai University, Tianjin 300071, China}

\author{Yu-xin Liu}\email{yxliu@pku.edu.cn}
\affiliation{Department of Physics and State Key Laboratory of Nuclear Physics and Technology, Peking University, Beijing 100871, China}
\affiliation{Collaborative Innovation Center of Quantum Matter, Beijing 100871, China}
\affiliation{Center for High Energy Physics, Peking University, Beijing 100871, China}

\author{Kh\'epani Raya}\email{khepani@nankai.edu.cn}
\affiliation{School of Physics, Nankai University, Tianjin 300071, China}

\date{\today}

\begin{abstract}
The mass relation ${M_{0^{+}}+3M_{1^{+\prime}}+5M_{2^{+}}= 9M_{1^{+}}}$ miraculously holds for the $P$-wave charmonium $(c\bar{c})$ and bottomonium $(b\bar{b})$ systems
with soaring precision. The origin of such relation can be addressed from Quark Models, and have been confirmed experimentally in a limited number of cases. In this connection, we propose $M_{0^{+}}+5M_{2^{+}}=3(M_{1^{+\prime}}+M_{1^{+}})$ as an extension to the $P$-wave $B_{c}$ case. In order to test its applicability, we employ a variety of Quark Model predictions for the $B_c$ mass spectrum. Our numerical analysis confirms
such formula is accurate up to very small deviations. 
\end{abstract}

%

\maketitle

\section{Introduction}
As it is well recognized, there are many puzzles and anomalies in hadron physics which have not received satisfactory explanations so far: \emph{e.g.} predicted states not yet observed, their computed masses, mass relations and decay patterns, etc. On a primary level, it hints that our knowledge on Quark Models (QM), which have played important roles in predicting and describing the hadron spectrum for decades~\cite{GellMann:1964nj,Zweig:1981pd,Godfrey:1985xj,Bijker:1994yr}, may possesses loopholes. Modern hadron physics suggests that the answers may emerge from our better understanding of non-perturbative Quantum Chromodynamics (QCD)~\cite{Fischer:2014cfa,Gomez-Rocha:2016cji,Chang:2019eob,Chen:2020ecu}. Nevertheless, simpler but reasonable explanations could come from an adequate recognition of hidden symmetries, which should determine the characters of the mass spectrum of hadrons and possibly
their production and decay patterns. Establishing identities between the mass spectra of hadrons with different quantum numbers may shed some light on the underlying symmetries which somehow remain hidden and being missed.


The $B_{c}$ states are quite special because they contain two heavy quarks with different flavors, thus existing in an intermediate mass regime between the charmonium and bottomonium systems, and maintaining those approximations that are sensible for the $c\bar{c}$ and $b\bar{b}$ sectors. At the same time, $B_c$ mesons are more stable than their equal-flavored counterparts, since they cannot annihilate into gluons. Hence, their study could provide a window to test our knowledge of hadron physics, specially concerning non-perturbative QCD. The ground-state pseudoscalar $B_{c}$ meson was first observed in 1998 by the CDF collaboration at the Tevatron~\cite{Abe:1998wi}.
More recently, two excited states of $B_{c}$ were reported by the CMS~\cite{Sirunyan:2019osb} and, later
confirmed, by the LHCb collaborations~\cite{Aaij:2019ldo}  at CERN; this success pushes the study on $B_{c}$ mesons forward.
Optimistically, with the joint effort of theorists and experimentalists, experiments at CERN provide an exceptional opportunity to
explore the charm-beauty ($B_{c}$) states and spectrum.

\section{Derivation and Analysis}
Experimentally, the masses of the spin-singlet $P$-wave states almost coincide with the spin-averaged centroid
of the triplet~\cite{Dobbs:2008ec,Adachi:2011ji} for the $c\bar{c}$ and $b\bar{b}$ systems.
The hyperfine mass splitting or the singlet-triplet mass splitting is almost zero~\cite{Titard:1994id,Lebed:2017yme,Peset:2018jkf}, hence one can safely write
 \begin{equation}\label{eq:hfsplitting}
 \Delta M_{hf} := \langle M(^3P)\rangle - M(^1P) = 0\;,
\end{equation}
with $ \langle M(^3P)\rangle = (M_{0^{++}} + 3M_{1^{++}} + 5M_{2^{++}})/9$ and $M(^1P) = M_{1^{+-}}$. This is further supported by the experimental data~\cite{Zyla:2020zbs}, from which: $ \langle M(^3P)\rangle/M(^1P) \approx 1.00002$ for $1P$ charmonium, while $\langle M(^3P)\rangle/M(^1P) \approx0.99994,\;0.99996$ for $1P$ and $2P$ bottomonium, respectively. So, amazingly, Eq.~\eqref{eq:hfsplitting} is well preserved. This observation indicates that even though the masses of $\chi_{c0}, \chi_{c1}$ and $\chi_{c2}$ are different, as we sum over the  masses of the corresponding $P$-wave states, with proper weights, the contributions of the spin-spin interaction among the states seem to be either negligible or mutually compensated with each other~\cite{Richard:2007wc}. In fact, in the context of QM, such weights are given by the spin-orbit term of the potential~\cite{Eichten:1979ms,Eichten:1978tg}.

Now, let us adapt the hyperfine mass splitting formula to the case of $B_c$ mesons. Firstly, for the unequal flavor systems the charge conjugation parity is no longer a good quantum number, but we can still define the spin-averaged mass of the corresponding triplet states as:
\begin{equation}\label{eq:hfsplittingB}
 \langle M(^3P)\rangle= (M_{0^{+}} + 3M_{^3P_1} + 5M_{2^{+}})/9\;, 
\end{equation}
where the singlet mass is $M(^1P) = M_{^1P_1}$. Note that, up to the first order of violating the unitary symmetry for the unequal flavor bound states,
the masses obey the so-called equal spacing rule (ESR)~\cite{Okubo:1961jc,GellMann:1962xb},
\begin{equation}\label{eq:srule0-}
 (M_{c\bar{c}} + M_{b\bar{b}})/2 = M_{c\bar{b}}\, .
\end{equation}
Under this assumption, it is clear that Eq.~(\ref{eq:hfsplitting}) still holds for these systems. Notice that the states with the same total angular momentum but different spins can mix, so the physical $|1^{+\prime} \rangle$ and $|1^+ \rangle$ are, in fact,  mixed states of $|^1P_1 \rangle$ and $|^3P_1 \rangle$. Thus, one has:
\begin{eqnarray}\label{eq:mixing}\nonumber
 |1^{+\prime} \rangle &=& \textmd{cos}\theta |^1P_1 \rangle + \textmd{sin}\theta |^3P_1 \rangle\, , \\
 |1^{+} \rangle &=& -\textmd{sin}\theta |^1P_1 \rangle + \textmd{cos}\theta |^3P_1 \rangle\, .
\end{eqnarray}
As the states \{$|1^{+\prime} \rangle$, $|1^{+} \rangle$\} and \{$|^1P_1 \rangle$, $|^3P_1 \rangle$\} are mutually related by a unitary transformation, the relation $M_{1^{\prime}} + M_{1^{+}} = M_{^1P_1} + M_{^3P_1}$ holds. Now, if we further assume $M_{^1P_1} \approx M_{^3P_1}$, which is valid so long as the meson contains only heavy quarks, then Eqs.~\eqref{eq:hfsplitting}-\eqref{eq:srule0-} yield:
\begin{eqnarray}
M_{0^{+}} + 5M_{2^{+}} &=& 3(M_{1^{+\prime}} + M_{1^{+}})\;.\label{eq:hfsplittingBc}
\end{eqnarray}
Naturally, the above expression has not been derived from first principles and some approximations have been made. Its formulation follows the assumption that Eq.~(\ref{eq:hfsplitting}) still holds for the $B_{c}$, given the associated ESR (experimentally~\cite{Zyla:2020zbs}, $\frac{M_{\eta_{c}}+M_{\eta_{b}}}{2M_{0^{-}}^{c\bar{b}}} \simeq 0.99$), and neglecting mild deviations induced by the mixing effects. These assumptions are valid due to the large masses of the systems in study. In fact, the spin and flavor symmetry breaking effects are suppressed by $1/(m_b m_c)$, the inverse of the heavy-quark masses~\cite{Neubert:2005mu}. Due to the lack of experimental measurements or lattice QCD results, we test validity of the proposed mass relation by employing QM~\cite{Godfrey:1985xj} predictions.  Specifically, we take the predicted spectrum from several QM approaches~\cite{Eichten:1994gt,Kiselev:1994rc,Gupta:1995ps,Fulcher:1998ka,
	Ebert:2002pp,Godfrey:2004ya,Li:2019tbn} and check whether or not (and, to what extent), the relation from Eq.~(\ref{eq:hfsplittingBc}) holds. For that purpose, we first define the deviation ratio:
\begin{equation}
\label{eq:ratio1}
r:=1+\delta_r=\frac{M_{0^{+}} + 5M_{2^{+}}}{3(M_{1^{+\prime}} + M_{1^{+}})}\;.
\end{equation}
Clearly, $\delta_r$ accounts for how much the above relation is being violated: while $\delta_r=0$ means Eq.~\eqref{eq:hfsplittingBc} holds perfectly, $\delta_r\approx 0$ supports the validity of the mass relation and our assumptions. The resulting values of  $r$ are listed in Table~\ref{tab:masscball}. Notably, for all listed QM calculations, the relative theoretical error lies below $0.2\%$ range, namely $|\delta_r| \textless 2\times 10^{-3}$. For example, taking the masses of the radial excited states, obtained recently in Ref.~\cite{Li:2019tbn}, we find that $r\simeq 1.001$. To give an idea, a value of $\delta_r\sim 10^{-3}$ implies a discrepancy, between the left-hand-side and the right-hand-side of Eq.~\eqref{eq:hfsplittingBc}, of order of $50$ MeV. The usual theoretical error associated with a quark
model is about 50 MeV.
\begin{table}[h!]
\caption{\label{tab:masscball} Masses of the $P$-wave $B_c$ mesons (in MeV), quoted from QM results~\cite{Eichten:1994gt,Kiselev:1994rc,Gupta:1995ps,Fulcher:1998ka,Ebert:2002pp,Godfrey:2004ya,Li:2019tbn}.}
\begin{tabular}{c|ccccccc}
\hline
$J^{P}$  &\cite{Eichten:1994gt} &\cite{Kiselev:1994rc}  &\cite{Gupta:1995ps} & \cite{Fulcher:1998ka} & \cite{Ebert:2002pp} 	&	\cite{Godfrey:2004ya} &	\cite{Li:2019tbn} \\
\hline
$2^{+}$            &6747 &6743 &6773  &6772  &6762   & 6768 & 6787  \\
$1^{+\prime}$  &6736 &6729 &6757  &6760  &6749   & 6750 & 6776  \\
$1^{+}$	       &6730 &6717  &6737  &6737 &6734   & 6741 & 6757  \\
$0^{+}$            &6700 &6683 &6688   &6701 &6699   & 6706 & 6714  \\
\hline
$r$            &1.0009 &1.0015 &1.0017   &1.0017 &1.0015   & 1.0018 & 1.0012  \\
\hline
\end{tabular}
\end{table}
\\
For a deeper, qualitative, understanding of the origin of $r\neq 1$, we can also follow the approach of J. Rosner~\cite{Rosner:1985dx}. There, for the $P$-wave mesons, the Hamiltonian in charge of the splitting and mixing of the multiplets is given by:
\begin{eqnarray}
\Delta H&=&c_{1}\vec{L}\cdot \vec{S}_{1}+c_{2}\vec{L}\cdot \vec{S}_{2}\nonumber\\&&+c_{T}\left(\frac{\vec{S}_{1}\cdot r\; \vec{S}_{2}\cdot r}{r^{2}}-\vec{S}_{1}\cdot \vec{S}_{2}\right)+c_{H}\vec{S}_{1}\cdot \vec{S}_{2}\;,\label{eq:HFinteraction}
\end{eqnarray}
thus the deviation ratio ($r$) can be expressed as
\begin{equation}
\label{eq:HFratio}
r=1+\frac{6 c_{H}}{4\bar{M}+c_{1}+c_{2}-c_{T}-5 c_{H}}\;.
\end{equation}
The above result relates deviations from $r=1$ with the spin-dependent terms, where $\vec{S}_{1}\cdot \vec{S}_{2}$ is expected to be zero (at leading order) for any $P$-wave state. Here, $\bar{M}$ corresponds to the eigenvalue of the spin-independent part of the Hamiltonian for $P$-states and the values $c_j$ are described in~\cite{Rosner:1985dx}. We also analyze how well Eq.~(\ref{eq:hfsplittingBc}) is preserved for  charmonium and bottomnium systems, where the charge conjugation states $(1^{+\prime},1^{+})\to( 1^{++},1^{+-})$ are taken into account. The empirical values~\cite{Zyla:2020zbs} produce $r\simeq 1$, with $\delta_r=4.12\times 10^{-3}$ for the charmonium and $\delta_r=7.2\times 10^{-4}$ for the bottomonium systems. 
%
%
%

Our simple analysis indicates that a mass relation for the $B_c$ systems is by no means trivial. Although some approximations were done along the derivation of Eq.~(\ref{eq:hfsplittingBc}), namely the equal spacing rule and taking $M_{^1P_1} \approx M_{^3P_1}$, those are practically harmless for the heavy systems. Some effects caused by the internal dynamics of the meson, could prevent Eq.~(\ref{eq:hfsplittingBc}) from being perfectly true, such as it is hinted by the hyperfine interaction part of the Hamiltonian, Eqs.~\eqref{eq:HFinteraction}-\eqref{eq:HFratio}. Nevertheless, our proposed equality could suggest a possible hidden symmetry or an underlying principle which induces the miraculous disappearance of the contribution of the spin-spin interaction when one sums up the masses of the $P_J$ states, with their proper weights. The disappearance of such term has been explained long ago with QM frameworks~\cite{Eichten:1978tg,Eichten:1979ms}, but the explanation from QCD's fundamental ingredients remains unclear. If we attribute this phenomenon to a conservation law, its deviation from $r=1$ is due to the quantum corrections to the Noether quantity.  It is also suggested that this deviation
arises from relativistic corrections and/or non-perturbative effects, which tend to affect more the light systems and become suppressed when moving towards the heavy sector~\cite{Chen:2018rwz}. This fact explains why the heavier the quark constituents are, the smaller the deviation from $r=1$  is. Amazingly, Eq.~(\ref{eq:hfsplitting}) still holds for positronium systems with very high precision~\cite{Lamm:2017lrn}, implying that non-relativistic approximations are still reasonable for this kind of bound states.  Also, to a lesser extent, the hyperfine splitting is small for some heavy-quark hybrid systems~\cite{Lebed:2017xih}. The latter makes it evident that non-perturbative effects, proper of the strong interactions in the Standard Model, are playing an important role.

\section{Summary}
Starting from the miraculous mass relation for the charmonium and bottomonium systems, given in Eq.~\eqref{eq:hfsplitting}, we derived the identity Eq.~\eqref{eq:hfsplittingBc} for the $B_c$ mesons. Subsequently, we employed a collection of QM results for the $B_c$ spectrum, and proved that the relation holds to a high precision. Also, our derivation implies that Eq.~\eqref{eq:hfsplitting} is still true for the $B_c$ systems, given the validity of the ESR. Similar outcomes have been also observed for positronium and heavy-quark hybrid systems~\cite{Lamm:2017lrn, Lebed:2017xih}, which can also be regarded as non-relativistic systems. The origin of Eq.~\eqref{eq:hfsplitting}, for heavy-quarkonia and positronium, can be addressed from QM: it is seen that crucial cancellations between the spin-dependent terms of the potential occur. This important outcome has been verified in a limited number of cases but, strongly suggesting that the intuition behind the QM potentials is correct, but predictions and explanations coming from QCD-connected studies could be beneficial and reveal something being missed. 

In the present work, we have seen that despite the fact of being an unequal flavor meson, splitting and mixing effects in the $B_c$ systems are still small, validating Eqs.~\eqref{eq:hfsplitting}-\eqref{eq:hfsplittingBc} for this case. This can also be understood from the Heavy Quark Effective Theory, in which the leading order Lagrangian explicitly shows spin-flavor symmetry, subsequently broken by the subleading terms, which are weighted by the inverse of the heavy-quark masses~\cite{Neubert:2005mu}. Furthermore, the spectroscopy of $B_c$ mesons could enlighten our understanding of exotic $B_c$-like structures~\cite{Ortega:2020uvc}. Those aspects should be further investigated. In particular, the mass relation for the $B_c$ systems that we have proposed must be tested by QCD-connected studies~\cite{Yin:2019bxe,Gutierrez-Guerrero:2019uwa,Chang:2019eob,Chen:2020ecu} and Lattice QCD simulations~\cite{Mathur:2018epb}, as well as future, more accurate, experimental measurements.

\section{Acknowledgements}
We acknowledge helpful conversations with Dr. Xian-hui Zhong.
This work is supported by: the Chinese Government Thousand Talents Plan
for Young Professionals and the National Natural Science Foundation of China
under contracts No. 11435001, and No. 11775041, the National Key
Basic Research Program of China under contract No. 2015CB856900.

\bibliographystyle{unsrt}
\bibliography{bibliography}
\end{document}